\documentclass[11pt, final, reqno, letterpaper]{article}

\usepackage[psamsfonts]{amssymb}
\usepackage[reqno]{amsmath}
\usepackage{latexsym,amsthm,setspace,amsfonts}
\usepackage{verbatim,enumerate,xspace}
\usepackage{hyperref, url}
\usepackage{epsfig}
\usepackage{color}
\usepackage{graphicx, booktabs, multirow}
\usepackage{natbib}
\usepackage{dsfont}
\usepackage[left=3cm, top=3cm, bottom=3cm, right=3cm]{geometry}

\graphicspath{{Figures/}}

\definecolor{orange}{rgb}{1,0.5,0}

\let\proglang=\textsf
\newcommand{\pkg}[1]{{\fontseries{b}\selectfont #1}}
\let\code=\texttt

\def\argmax{\mathop{\rm argmax}}

\def\bi{\begin{itemize}}
\def\ei{\end{itemize}}
\def\bea{\begin{eqnarray*}}
\def\eea{\end{eqnarray*}}
\def\be{\begin{equation}}
\def\ee{\end{equation}}
\def\bean{\begin{eqnarray}}
\def\eean{\end{eqnarray}}
\def\bd{\begin{description}}
\def\ed{\end{description}}
\def\bl{\begin{list}{$\bullet$}{}}   
\def\cl{\begin{list}{$\circ$}{}}     
\def\el{\end{list}}                  

\def\b1{\mathbf{1}}

\newcommand{\CC}{\mathcal{C}}

\newcommand{\CCm}{{\CC_m}}
\newcommand{\II}{\mathcal{I}}
\newcommand{\KK}{\mathcal{K}}
\newcommand{\RR}{\mathbb{R}}
\newcommand{\UU}{\mathbb{U}}
\newcommand{\FF}{{\mathbb F}}
\newcommand{\YY}{{\mathbb Y}}
\newcommand{\WW}{{\mathbb W}}
\newcommand{\HH}{{\mathbb H}}

\newcommand{\ZZ}{{\mathbb Z}}

\newcommand{\mle}{\widehat p}
\newcommand{\lmle}{\widehat \psi}
\newcommand{\emp}{\bar{p}}
\newcommand{\Hhat}{\widehat{{\mathbb H}}}

\newcommand{\What}{\widehat{{\mathbb W}}}
\newcommand{\mc}[1]{\mathcal {#1}}

\newcommand{\kl}[2]{\rho_{\mathrm{KL}}\left({#1} \, \Vert \, {#2} \right)}
\newcommand{\argmin}{\mathrm{argmin}}

\theoremstyle{plain}
\newtheorem{theorem}{Theorem}[section]
\newtheorem{corollary}[theorem]{Corollary}

\newtheorem{lemma}[theorem]{Lemma}
\newtheorem{proposition}[theorem]{Proposition}

\newtheorem{definition}{Definition}[section]

\newtheorem{remark}{Remark}[section]

\numberwithin{equation}{section}

\title{Asymptotics of the discrete log-concave maximum likelihood estimator and related applications}

\author{Fadoua Balabdaoui$^1$,
Hanna Jankowski$^2$,
Kaspar Rufibach$^3$, \\ and Marios Pavlides$^4$}

\date{September 30, 2012}

\begin{document}

\maketitle

\begin{center}
$^{1}$CEREMADE, Universit\'e Paris-Dauphine, Paris, France \\
$^{2}$Department of Mathematics and Statistics, York University, Toronto, Canada\\
$^{3}$Division of Biostatistics, Institute for Social and Preventive Medicine, University of Zurich, Zurich, Switzerland\\
$^{4}$Centre for Statistical Science and Operational Research, Queen's University Belfast, Belfast, Northern Ireland, United Kingdom\\

\end{center}
\begin{abstract}
The assumption of log-concavity is a flexible and appealing nonparametric shape constraint in distribution modelling.  In this work, we study the log-concave maximum likelihood estimator (MLE) of a probability mass function (pmf).  We show that the MLE is strongly consistent and derive its pointwise asymptotic theory under both the well-- and misspecified settings.  Our asymptotic results are used to calculate confidence intervals for the true log-concave pmf.   Both the MLE and the associated confidence intervals may be easily computed using the \proglang{R} package \pkg{logcondiscr}.  We illustrate our theoretical results using recent data from the H1N1 pandemic in Ontario, Canada.
\end{abstract}

\bigskip


\section{Introduction}\label{S: intro}

Nonparametric maximum likelihood estimation of a log--concave probability density in the continuous setting has attracted considerable attention over the last few years.
The list of references is extensive, and we refer the reader to \citet{walther_08, cule_08, seregin_10} and the references therein for an overview of recent theoretical and computational developments.    The merits of using log--concavity as a shape constraint have been discussed in detail in \cite{balabdaoui_09}, \cite{cule_08}, \cite{walther_08}, and \cite{duembgen_logcon10} for the continuous setting.    

Given the large corpus of work on estimation of a log--concave density in the continuous case, it comes as a surprise that little attention has been given to estimation of a log--concave probability mass function (pmf).    The log-concave assumption provides a broad and flexible, yet natural, non-parametric class of distributions on $\ZZ,$ and many popular discrete parametric models admit a log-concave pmf.   Binomial, negative binomial, geometric, hypergeometric, uniform, Poisson, hyper-Poisson \citep{bardwell_64, crow_65}, the P\'{o}lya-Eggenberger, and the Skellam distribution \citep{karlis_06, alzaid_10} are some examples; see \cite{MR0268996} and \cite{devroye_87} for further details.  In Section \ref{disclogconc}, we discuss more thoroughly the  properties and benefits of the class of log-concave pmfs on $\ZZ.$ 

The unpublished Master's thesis of \cite{weyermann_08} is the only previous work on the MLE of a log-concave pmf of which we are aware. In \citet{weyermann_08}, it was shown that the MLE of a log-concave pmf on $\ZZ$ exists and is unique.   To compute the MLE,  \cite{weyermann_08} provided an active set algorithm (implemented in \proglang{Matlab}), much in the spirit of \cite{duembgen_07}.  We have adapted this code to \proglang{R} \citep{R} in the new package \pkg{logcondiscr} \citep{logcondiscr}, available from CRAN.  

In this work, we study consistency and asymptotic properties of the log-concave MLE, including the setting when the model has been misspecified.   In Section \ref{MLE}, we recall some of the main results of \citet{weyermann_08}, and provide additional characterisations of the estimator.  These characterisations are important as they provide insight into its asymptotic behaviour.   In this section, we also establish consistency of the MLE.    When the true pmf $p_0$ is log-concave, the MLE converges to $p_0.$  However, if the model is misspecified, the MLE converges to the log-concave pmf which is closest to $p_0$ in the Kullback-Leibler divergence.  We denote this pmf as $\widehat p_0,$ and refer to it as the Kullback-Leibler (KL) projection of $p_0$.  Similar results have also been shown in the continuous setting \citep{cule_08, cule_10_ejs}.

In Section \ref{Asym}, we study the asymptotic behaviour of the log-concave MLE, $\widehat p_n,$ in the well- and misspecified setting.  Section \ref{sec:tight} gives some preliminary results on tightness of the estimator, while Section \ref{sec:pointwise}  {provides} the limiting behaviour of the MLE.  Our main results establish the pointwise asymptotic distribution of $\sqrt{n}(\widehat p_n - \widehat p_0),$  and the limiting distribution is explicitly given.  This limit may be characterised in terms of an envelope-type process, $\mathbb H$, which can be viewed as a discrete analogue of the random process of \cite{balabdaoui_09}, also appearing in the pointwise convergence of the MLE of a convex decreasing density.  This type of process was first described in the pioneering work of \cite{gjw01A, gjw01B}.
In the study of the asymptotic distribution at a point $x,$  we need to make certain technical assumptions:  In the well-specified setting, we assume that $p_0$ has one-sided support, and when the model is misspecified, we assume that the true pmf has bounded support.  {Moreover,  we assume in the well-specified setting that $x$ does not} lie in a region where $\log (p_0)$ is linear on an infinite subset of $\ZZ.$ This excludes distributions such as the geometric from our {analysis}.   Our results also show that if $\log(p_0)$ is strictly concave, then the log-concave MLE will have the same limiting distribution as the empirical pmf ({similar results have been proved} for the Grenander estimator of a pmf;  see \citealp{jankowski_09}).  {For small sample sizes, however, our simulation study in Section \ref{sec:finite}} shows that the behaviour of the log-concave MLE can be significantly better than that of the empirical pmf.

In both the well- and misspecified case, we show that the limiting process can be described as the solution of a least squares concave regression problem.   In the well-specified case, this solution can be found explicitly using the \proglang{R} package \pkg{cobs} \citep{cobs}.   Therefore, we are able to sample directly from the limiting distribution, which allows us to compute pointwise confidence intervals for the true pmf when it is log-concave.  The details of our approach are given in Section \ref{CIs}, and the method is implemented in the \proglang{R} package \pkg{logcondiscr}.

For the MLE of a monotone density,   \cite{patilea} gives rates of convergence in the misspecified setting in a modified Hellinger distance.  Some additional results are proved in \citet{jankowski_gren}. In the present paper, we go beyond convergence rates by explicitly giving the limiting distribution of the MLE under misspecification.  We also believe that this is the first work where confidence intervals have been explicitly computed in the log-concave (well-specified) setting.   In the continuous case, \citet{balabdaoui_09} derive pointwise asymptotics for the MLE of a continuous log-concave density $f$.  However, these depend on the value of $\psi''(x_0)$, where $\psi=\log f,$ which is difficult to estimate.  Similar problems arise for the monotone shape-restriction, and the work of {\cite{banerjwell01}} was designed to overcome this issue.  Currently, no such results exist for the MLE of log-concave density $f$ on $\RR$.

As an illustration of our {methods}, we apply the proposed estimator to a real data set of H1N1 influenza pandemic data from Ontario, Canada.   This data comes from an early study of the pandemic, \citet{tuite_10}, when it was important to provide a quick analysis of the behaviour of the virus.  The flexibility of the log--concave assumption makes it suitable to describe important aspects of the incubation period of the swine flu, as well as the duration of symptoms.  To handle {potential} errors in the data collection, we use a simple mixture model, which gives even more flexibility to our {approach}.  This mixture model is also implemented in the \proglang{R} package \pkg{logcondiscr}.

Conclusions and a discussion can be found in Section \ref{conclusion}. All proofs and additional technical details can be found in the technical report \cite{BJRProofs2012}.

\section{Log--concavity of discrete distributions}\label{disclogconc}

The class of log-concave distributions is a natural assumption to make in practice,  and is particularly popular in economics, see for example \citet{an, MR2213177}, who consider both the continuous and discrete settings.    Additional references on log-concavity in the discrete setting include \citet{KG1971} and \citet{dharma88}.

For a discrete random variable $X$ with state space contained in the integers $\ZZ,$ we define the probability mass function
$p(z)=P(X=z)$ for $z\in \ZZ.$  That is, $p:\ZZ \mapsto [0,1]$ such that $\sum_{z\in \ZZ} p(z)=1.$  We denote the support of $p$ as $\mathcal S = \{z \in \ZZ: p(z) > 0 \}$.


\begin{definition}
A pmf $p$ with support $\mathcal S\subset \ZZ$ is log-concave if both of the following conditions hold
\begin{list}{--}
        {\setlength{\topsep}{3pt}
        \setlength{\parskip}{0pt}
        \setlength{\partopsep}{0pt}
        \setlength{\parsep}{0pt}
        \setlength{\itemsep}{3pt}
        \setlength{\leftmargin}{10pt}}
\item If $z_1<z_2<z_3$ are integers such that $p(z_1)p(z_3)>0$, then $p(z_2)>0.$
\item $p(z)^2\geq p(z-1)p(z+1)$ for all $z\in \ZZ.$
\end{list}
\end{definition}
For a pmf $p$, let $\psi(z) = \log p(z),$ and $[\Delta \psi](z)=\psi(z+1)-2\psi(z) + \psi(z-1)$ denote the discrete Laplacian of $\psi$ . The following is an equivalent definition of log-concavity.

\begin{proposition}\label{EquivCond2}
A pmf $p$ with support $\mathcal S\subset \ZZ$ is log-concave if and only if $\mathcal S$ is a connected subset of $\ZZ$ and
$[\Delta \psi](z) \le 0
$ for all $z \in \mathcal S$.
\end{proposition}

In the discrete setting, the class of log-concave distributions has numerous appealing attributes.   As noted in \citet{devroye_87}, this class of distributions is ``vast" in that it includes many of the classical discrete models.    This allows one to specify a class of distributions instead of a single parametric family, greatly increasing the robustness of the results at a surprisingly modest loss of efficiency.   Many properties of discrete log-concave distributions were identified in \citet{an}:   Log-concavity is preserved under convolutions, truncation, and increasing transformations.   A log-concave pmf $p$ is necessarily unimodal, but it need not be symmetric.  Moreover, the mode need not be specified a priori.  The log-concave pmf also has a relatively nice tail behaviour, in that it has at most a geometric tail and always admits a moment generating function.

Although unimodality is one of its identifying features, the class of log-concave pmfs is smaller than the class of unimodal pmfs.   Log-concave distributions are unimodal, but not all unimodal distributions are log-concave.  In fact, a distribution is log-concave if and only if it is \emph{strongly unimodal}  (a pmf $p$ is said to be strongly unimodal if, for any unimodal pmf $q$, the convolution $p\star q$ is also unimodal, cf. \citealp{Ibragimov56}).   Furthermore, as Proposition \ref{EquivCond2} shows, a log-concave pmf must have a concave logarithm, whereas $\log(p)$ only needs to be unimodal for a unimodal pmf $p$.  Therefore, we can think of log-concave pmfs as more ``smooth" than unimodal ones.   When choosing a nonparametric class, one hopes to pick a class that is both ``large" and ``small" at the same time.  That is, one would want the class to be sufficiently large that it encompass most potential distributions of interest.  On the other hand, the smaller the chosen class, the greater the improvement in estimation accuracy compared to a purely nonparametric estimate.   It is our view that the class of log-concave pmfs achieves a good balance between accuracy and robustness.

In the continuous setting, the maximum likelihood estimator of a unimodal density does not exist, which makes the log-concave assumption one natural substitute for the unimodal setting.  However, the appeal of the log-concave assumption is much greater.  We share the view of \citet[Section 2]{cule_08} that the ``class of log-concave densities is a natural, infinite dimensional generalization of the class of Gaussian densities".    Although the discrete Gaussian model is not prevalent in the statistics literature, we feel that this statement of \cite{cule_08} continues to hold in the discrete setup, in that the class of log-concave distributions is a very natural, yet flexible, class of pmfs.

Log-concavity of a pmf can also be described through the following alternative definition, connecting log-concavity and monotonicity.

\begin{proposition}\label{EquivCond22}
A pmf $p$ with support $\mathcal S $ is log-concave if and only if $\mathcal S$ is a connected set of integers  and the sequence $\{ p(z)/p(z-1), z \in \mathcal S \}$ is nonincreasing.
\end{proposition}

Note that if $p$ has one-sided support of the form $[z_0, \infty) \cap \ZZ$ for some $z_0 \in \ZZ$, the first ratio term takes the value $\infty$. The proposition clearly gives the possibility of constructing an estimator based on \lq\lq monotonising\rq\rq \ the empirical probability ratios. This alternative approach will be pursued elsewhere.


Modelling data via a discrete distribution is quite natural in many applications.  One such example is the case when the observed data have been grouped, as in the H1N1 example considered in Section~\ref{Examples}, or discretised.   That is, let $\{A_z, z\in\ZZ\}$ denote a partition of the positive real line such that $|A_z|$ is constant as the index $z$ varies.  We assume that each interval $A_z$ is either of the form $(\alpha_z, \beta_z]$ or $[\alpha_z, \beta_z).$   For a continuous random variable $X$ we then define the probability mass function $p$ as $p(z)=P(X\in A_z)$.  Such a scenario arises if one observes $Y=\delta\lfloor X/\delta\rfloor,$ for example, instead of the continuous random variable $X.$

\begin{proposition}\label{prop:grouping}
Suppose that the continuous random variable $X$ has a log-concave density with respect to Lebesgue measure on $\RR.$  Then the probability mass function  $p(z)=P(X\in A_z)$ is also log-concave.
\end{proposition}
Throughout this paper, we focus on the probability mass function defined on $\ZZ$.  However, our results are applicable to a pmf defined on any regular grid, as long as that grid does not depend on the sample size, in contrast to what was considered in {\cite{mouli11}}.


\subsection{The Kullback-Leibler projection}

\begin{figure}[!h]
\begin{center}
\centerline{\includegraphics[angle=-90, width=0.8\textwidth]{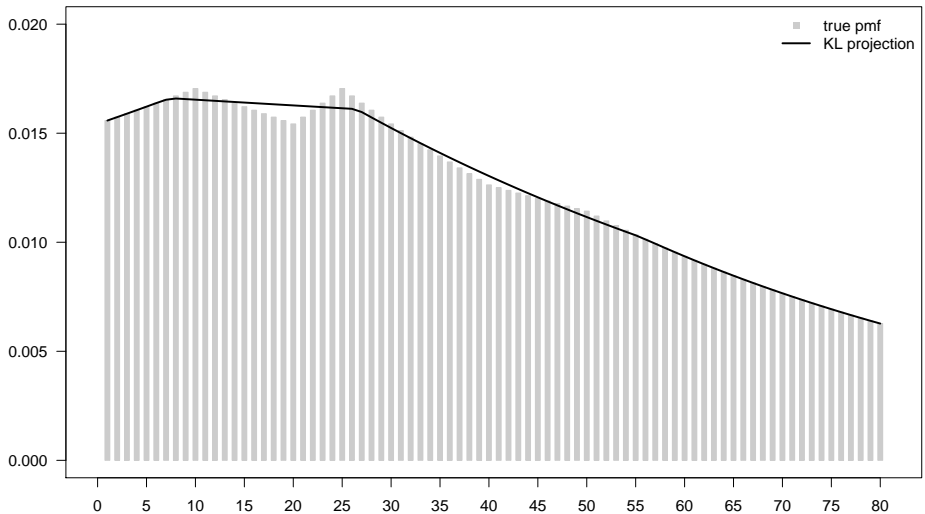}}
\end{center}
\vspace*{-1cm}
\caption{\small An example showing $p_0$ (with support  $\{1, \ldots, 80\}$) and its Kullback-Leibler projection~$\widehat p_0.$
} \label{fig:KLprojection}
\end{figure}

Next, fix a probability mass function $p_0$ and write $\psi_0 = \log p_0.$  For a pmf $p$ on $\ZZ,$ we define
\begin{eqnarray*}
\kl{p}{p_0} &=& \sum_{z \in \ZZ} \log \left(\frac{p_{0}(z)}{p(z)}\right) p_{0}(z)
\end{eqnarray*}
to denote the Kullback-Leibler (KL) divergence of $p$ from $p_0$.  Let $\mc{LC}_1$ denote the class of log-concave pmfs on $\ZZ$.  
The following theorem gives existence and uniqueness of the KL projection of $p_0$ on the class $\mc{LC}_1$. The result can be viewed as a discrete version of \citet[Theorem 4]{cule_10_ejs}.

\medskip

\begin{theorem}\label{KLproj}
Suppose that $p_0$ is a discrete probability mass function on $\ZZ$ with finite mean such that $|\sum_z  p_{0}(z) \log p_{0}(z)| < \infty.$  Then there exists a unique log-concave pmf on $\ZZ$, $\widehat p_0,$ such that
$\widehat p_0 = \argmin_{p\in \mc{LC}_1} \kl{p}{p_0}.$
\end{theorem}

Figure \ref{fig:KLprojection} shows an example of a non log-concave $p_0$ and the associated KL projection, $\widehat p_0$, computed using the package \pkg{logcondiscr}.

\begin{definition}\label{def:knot}
Let $\varphi$ denote a concave function on $\ZZ$ such that  $\varphi(z) < \infty$ for all $z \in \ZZ$. A point $x\in \{\ZZ:  \varphi(z)>-\infty\}$ is a \emph{knot} of $\varphi$ if $\varphi$ changes slope at $x$; i.e., $[\Delta \varphi](x) < 0$.   A point $x$ is called a \emph{double knot} of $\varphi$ if both $x$ and $x+1$ are knots.  A point $x$ is called a \emph{triple knot} of $\varphi$ if $x-1, x,$ and $x+1$ are knots.    A point $x$ is called an \emph{internal knot} of $\varphi$ if $x$ is knot of $\varphi$ and $\varphi(x-1), \varphi(x+1)>-\infty.$
\end{definition}

\medskip

The next lemma gives a characterization of the log-concave KL projection of $p_0$.

\begin{lemma}\label{lem:KLprojection}
Suppose that $p_0$ satisfies the conditions of Theorem \ref{KLproj}.  Then $\widehat p_0$ is the Kullback-Leibler projection of $p_0$ if and only if it satisfies  $\sum_{z=-\infty}^\infty \widehat p_{0}(z) =1$ and
\begin{eqnarray*}
\sum_{z=-\infty}^{x-1} F_0(z)
\left\{ \begin{array}{ll}
\geq & \sum_{z=-\infty}^{x-1} \widehat F_0(z),  \ \forall \ x \in \ZZ \\
 = & \sum_{z=-\infty}^{x - 1} \widehat F_0(z),  \ \textrm{if $x$ is a knot of $\widehat \psi_0$}
 \end{array}\right.
\end{eqnarray*}
where $F_0$ and $\widehat F_0$ are the cumulative distribution functions based on $p_0$ and $\widehat p_0,$ respectively.
\end{lemma}

\citet{duembgen_10} study the KL projection for the continuous density on $\RR^d.$  They provide a similar characterization to that above for the case $d=1,$ along with some additional properties of the KL projection.   Such properties could also be derived for the discrete case, with appropriate modifications, although we do not pursue these here.    In the discrete case,  the package \pkg{logcondiscr} may be used to calculate $\widehat p_0$ directly, at least whenever $p_0$ has a bounded support.

\section{Properties of the maximum likelihood estimator}\label{MLE}
Let $(X_1, \ldots, X_n)$ denote a random sample from the pmf $p_0$ where $n \ge 3$.  Then the MLE of a log-concave pmf is found by maximising the log-likelihood $\sum_{i=1}^n \log p({X_i}) /n$ over $\mathcal{LC}_1.$    Let ${\emp_{n}(z)} = n^{-1}\sum_{i=1}^n 1_{\{X_i = z\}}$, denote the empirical pmf.   By Theorem 3.1 of \cite{silverman82}, the MLE can be found as the maximiser of
\begin{eqnarray*}
\sum_{z \in \ZZ} \bar p_{n}(z) \log(p(z)) - \sum_{z \in \ZZ} p(z)
\end{eqnarray*}
over the class $\mathcal{LC}$ (the class of log-concave nonnegative sequences).   Equivalently, the MLE exists if and only if the criterion function
\begin{eqnarray*}
\label{Phin}
\Phi_n(\psi) &= & \sum_{z \in \ZZ} \bar{p}_{n}(z) \psi(z) -  \sum_{z \in \ZZ} \exp \psi(z)
\end{eqnarray*}
admits a maximiser {$\lmle_n$} over $\mathcal C$ (the class of all concave functions).  Then, the maximum likelihood estimator $\mle_n$ is given by  {$\mle_{n}(z) = \exp \lmle_{n}(z)$} for $z \in \ZZ$.

Reducing the set of functions over which $\Phi_n$ is maximised is one of the key steps in proving existence of {$\lmle_n$}. It also sheds more light on the shape of the estimator, and is of crucial importance when setting up an algorithm to compute {$\mle_n$}. Let $m$ be the number of distinct values in $(X_1, \ldots, X_n)$, and let $\II = \{z_1, \ldots, z_m\}$ denote the set of unique values in the sample in $(X_1, \ldots, X_n)$.   We also order the values in $\II$ so that  $z_1 <\ldots < z_m.$ Define the family of functions
\bea
    \mathcal{F}_m &:=& \{\varphi \ : \ \ZZ \to [-\infty, \infty), \ \varphi = -\infty \text{ on } \ZZ \cap \{\RR \setminus [z_1, z_m]\}\}.
\eea
For any $\varphi \in \mathcal{F}_m$, we consider the set of knots  $\KK(\varphi) = \{k \in \ZZ \cap [z_1, z_m] \ : \ [\Delta \varphi](k) <0 \}.$  {Note that $z_1$ and $z_m$ are always} in $\KK(\varphi)$.
Finally, we consider the sub-family
$ \mathcal{F}_m(\II) = \{\varphi \in \mathcal{F}_m \ : \ \KK(\varphi) \subseteq \II  \} $
of functions in $\mathcal{F}_m$ which only admit knots in the set of observations, 
and we let $\mathcal{C}_m(\II)$ be the subset of concave functions $\varphi$ in $\mathcal{F}_m(\II)$.

\begin{theorem}[\citealp{weyermann_08}]\label{theo: existence}
Maximisation of $\Phi_n$ over $\cal C$ is equivalent to its maximisation over $\CCm(\II)$.  Furthermore,
the maximiser
\bea
{\lmle_n} &:=& \argmax_{\varphi \in \CCm(\II)} \Phi_n(\varphi)
\eea
exists and is unique.
\end{theorem}


Therefore, attention can be restricted to concave functions $\varphi$ such that $\varphi = -\infty$ outside $[z_1, z_m] \cap \ZZ$ and having knots only in the set of observations. If $k_1, \ldots, k_p$ denote the internal knots of $\lmle_n$, then it is not difficult to see that $\lmle_n$ must have the following form
\bean\label{Formpsi}
\lmle_{n}(z) = a + b z + \sum_{i=1}^p c_i (k_i - z)_+, \ \ z \in \ZZ \cap [z_1, z_m]
\eean
where $a, b \in \RR$ and $c_i < 0$.  Here, we have used the standard notation $z_+ = z 1_{\{z\geq0\}}.$ 

\begin{remark}\label{remark:DOF}
Given the location of the knots as in \eqref{Formpsi}, to find the MLE one needs only to find the $p+2$ unknown values of $a,b, c_1, \ldots, c_p.$  
From Lemma \ref{CharMLE} below, we know that the MLE satisfies $p+1$ equalities in \eqref{FenchelEq}, plus $\sum_{x} \mle_{n}(x)=1.$  Hence, we have $p+2$ equations with $p+2$ unknowns, as long as the locations of the knots are known.   In essence, this tells us that the ``degrees of freedom" of the estimator is equal to  the number of knots.  We believe that this characteristic is one key to the quality of the performance of the MLE, as compared to, for example, the empirical estimator, which has more degrees of freedom.    We shall make use of this heuristic when we develop our confidence intervals in Section~\ref{CIs}.
\end{remark}

In the study of shape-constrained estimators, characterisations provide invaluable insight into their behaviour.   These are often referred to as the Fenchel conditions, due to their relationship with Fenchel duality in convex optimization problems.  The characterisation of the MLE of a log-concave pmf is given below.  Note that it shares a lot of similarity with the characterisation in the continuous setting \citep[Theorem 2.4]{duembgen_09}.   In what follows, $\FF_n$ denotes the empirical cumulative distribution function of the sample $(X_1, \ldots, X_n)$.

\begin{lemma}\label{CharMLE}
Let $\tilde \psi \in \mathcal{C}_m(\mathcal I)$ such that $\tilde F_n(y) = \sum_{z=z_1}^y \exp \tilde \psi(z), \ y \in \ZZ \cap [z_1, z_m]$
satisfies $\tilde F_n(z_m) = 1$. Then,  $\tilde \psi  = \lmle_n$ if and only if the following conditions hold
\bean
\sum_{z=z_1}^{x-1} \mathbb  F_n(z)
\left\{ \begin{array}{ll}
\geq & \sum_{z=z_1}^{x-1} \tilde F_n(z),  \ \forall \ x \in \ZZ \cap [z_1, z_m] \label{FenchelIneq} \ \ \\\\
 = & \sum_{z=z_1}^{x - 1} \tilde F_n(z),  \ \textrm{if $x$ is a knot of $\tilde \psi$.}  \ \ \label{FenchelEq}
\end{array}\right.
\eean
We use the convention that both sums are equal to 0 if $x = z_1$.
\end{lemma}

\bigskip

For double and triple knots, the characterization above reduces to simple forms as we show in the following corollary.

\medskip
\begin{corollary}\label{cor:double}
Suppose that $k$ is a double knot of the log-MLE.  Then $\widehat F_n(k)=\FF_n(k).$  If $k$ is a triple knot of the log-MLE, then $\widehat p_{n}(k)=\emp_{n}(k).$
\end{corollary}

Using similar techniques to those used to prove Lemma \ref{CharMLE}, further properties of the MLE can be established \citep[Proposition C.1]{BJRProofs2012}.  For instance, it can be shown that
\begin{eqnarray}
\sum_{z} z \mle_{n}(z) &=& \sum_z z \emp_{n}(z) \label{line:EqMean} \\
\sum_{z} |z-a|^m \mle_{n}(z) &\leq & \sum_z |z-a|^m \emp_{n}(z)\label{line:ExpBound}
\end{eqnarray}
for any $a\in \RR$ and $m\geq 1.$ Hence, the MLE has the same mean as the empirical distribution and a smaller variance than the empirical distribution.   Similar bounds were observed in \citet{duembgen_09} and \citet[Remark 2.3]{duembgen_10} for the MLE of a continuous log--concave density.


\subsection{Consistency}

For two probability mass functions $p$ and $q,$ let $\ell_k(p, q)$ denote the distance $\left(\sum_{z \in \ZZ} (p(z) - q(z))^k\right)^{1/k}$ if $1 \le k < \infty $, and $\sup_{z \in \ZZ}  \vert p(z) - q(z) \vert$ if $k = \infty$.  Also, let $h^2(p,q)=2^{-1} \sum_{z\in \ZZ} (\sqrt{p(z)} - \sqrt{q(z)})^2$ denote the Hellinger distance.  The next statement gives conditions under which consistency is observed.   These conditions (as well as the proof of the result) are similar to that of \citet{cule_10_ejs}.  The theorem also provides an alternative way of showing existence and uniqueness of the MLE.

\begin{theorem}\label{ConsistencyMLE}
Suppose that $p_0$ is a discrete distribution on $\ZZ$ with finite mean such that $|\sum_z  p_{0}(z) \log p_{0}(z)| < \infty.$  Then $d(\mle_n, \widehat p_0)  \rightarrow 0$ almost surely, where $d$ is the distance $\ell_k$ for any $1\leq k \leq \infty$ or the Hellinger distance $h.$
\end{theorem}

Thus, even if the class $\mc{LC}_1$ was originally misspecified, the MLE converges to the pmf which is closest, in the KL divergence, to the true pmf. Of course, if $p_0$ is log-concave, then $\widehat p_0=p_0,$ and our result implies that the MLE is consistent (note that any log--concave pmf satisfies the conditions of the theorem).   The result also implies consistency of cumulative distribution functions.

\begin{corollary}\label{cor:glivenko}
Let $\widehat F_n (y)= \sum_{z \leq y}  \mle_{n}(z),$ and let $\widehat F_0(y) = \sum_{z \leq y} \widehat p_{0}(z).$  Then, under the conditions of Theorem \ref{ConsistencyMLE},  $\sup_{y\in \ZZ} |\widehat F_n(y)- \widehat F_0(y)| \rightarrow 0$ almost surely.
\end{corollary}

Let $\widehat \psi_0=\log (\widehat p_0).$   The following result states that knots of $\lmle_n$ are also consistent estimates of the knots of $\widehat \psi_0$.

\begin{lemma}\label{lemma:knots}
For any knot point $r$ of $\widehat \psi_0$, there exists a positive integer $n_0$ sufficiently large such that for all $n\geq n_0,$ $r$ is also a knot of the MLE $\lmle_n$ with probability one.
\end{lemma}

From a technical point of view, Lemma \ref{lemma:knots} is crucial for deriving weak convergence of our estimator. In practice, it implies that a knot of $\widehat \psi_0$ is also a knot of the log-MLE $\lmle_n $ when the sample size is large enough. The same lemma does not say anything about the converse property, and an observed knot of $\lmle_n$ is not necessarily a true knot.   The confidence intervals derived in this work rely, however, on our knowledge of the knot points.  What allows us to overcome this issue is Remark \ref{remark:DOF}:  Namely, assuming more knots than necessary only increases our degrees of freedom.  In Section \ref{CIs} we discuss further the impact of the knot points on confidence intervals.

\vspace{-0.2cm}

\subsection{Finite sample behaviour}\label{sec:finite}

\begin{figure}[h!]
\begin{center}
\centerline{\includegraphics[width = 0.5\textwidth]{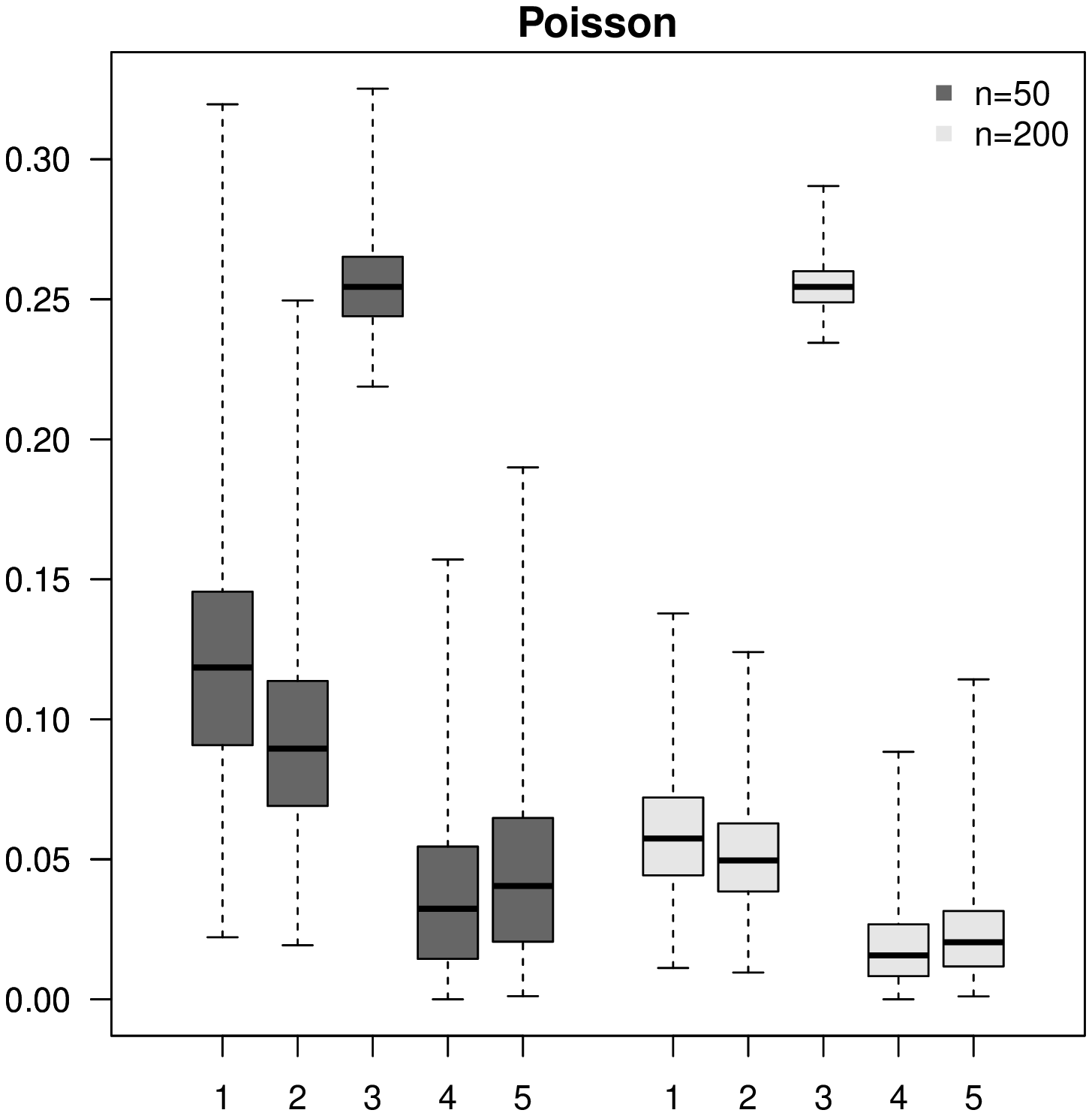}
\includegraphics[width = 0.5\textwidth]{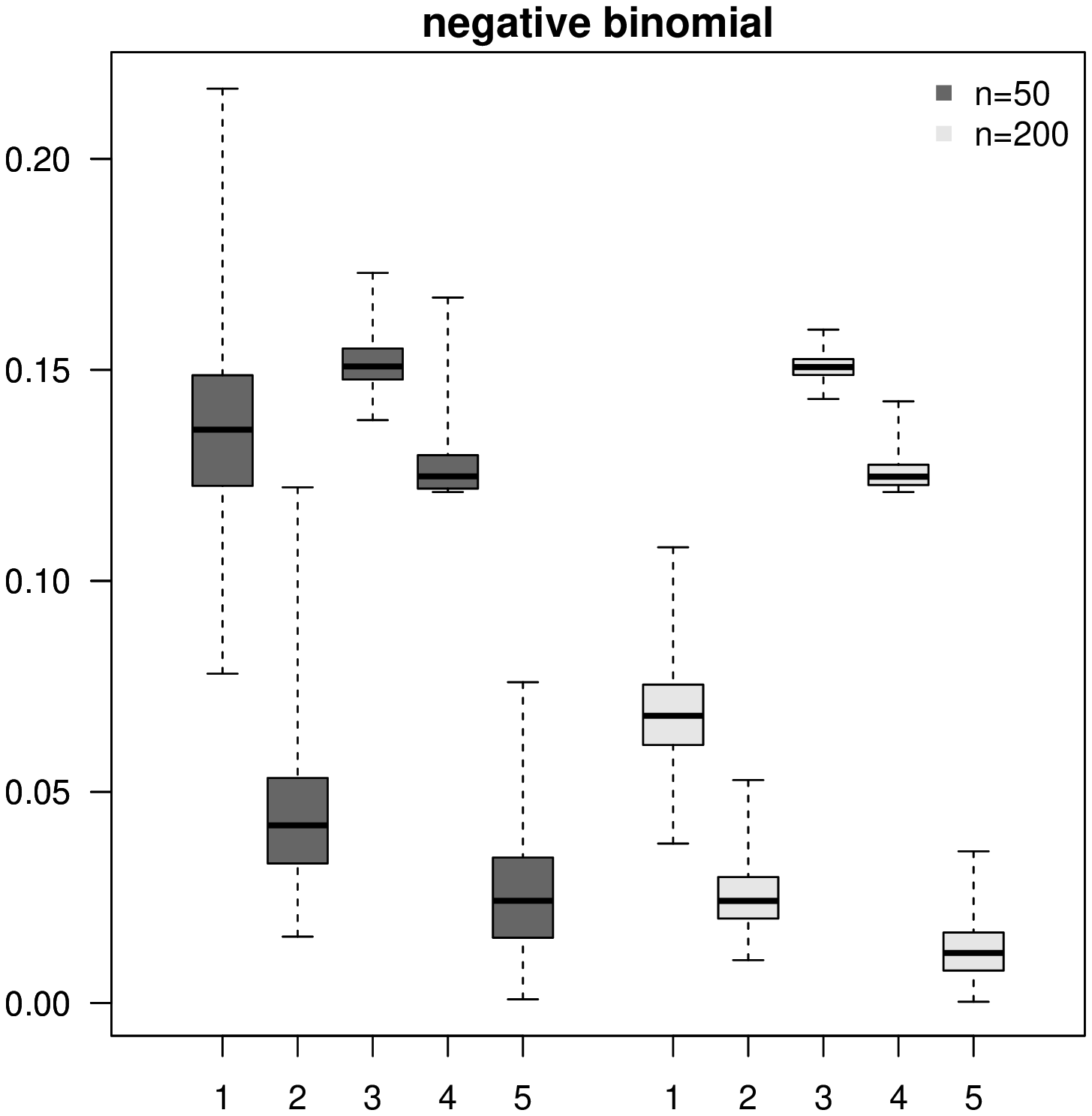}}
\end{center}
\vspace{-0.5cm}
\caption{Boxplots of the $\ell_2$ distance of the estimated pmf from the true pmf under each of the cases ($1$--$5$) listed in the text.  Each box plot is the result of $B=1000$ simulations.  On the left the true distribution is the Poisson ($\lambda=2$) and on the right it is the negative binomial ($r=6$, $p=0.3$). }
\label{fig:smallsample}
\end{figure}

To learn about the behaviour of the estimator for finite sample sizes, we compare the results of several nonparametric and parametric maximum likelihood estimators when sampling from the Poisson ($\lambda=2$) and negative binomial ($r=6$, $p=0.3$) distributions.  In each case we calculate the following:

\begin{list}{}
        {\setlength{\topsep}{3pt}
        \setlength{\parskip}{3pt}
        \setlength{\partopsep}{0pt}
        \setlength{\parsep}{0pt}
        \setlength{\itemsep}{3pt}
        \setlength{\leftmargin}{10pt}}
\item 1.  The empirical pmf (the MLE with no underlying assumptions).
\item 2.  The log-concave MLE.
\item 3.  The parametric MLE assuming the geometric distribution.
\item 4.  The parametric MLE assuming the Poisson distribution.
\item 5.  The parametric MLE assuming the negative binomial distribution.
\end{list}
Our results are shown in Figure \ref{fig:smallsample}, where we compare the $\ell_2$ distance of the true pmf to the estimated pmf in each of these cases.  The power of the log-concave assumption is clearly shown in these simulations.  The log-concave MLE performs well in estimating both distributions, albeit not as well as the correct parametric MLE.  Making the incorrect assumption carries with it the greatest cost.  Note, however, that the negative binomial MLE performs well for the Poisson distribution.  This is because the negative binomial converges to the Poisson when $p=\lambda/(\lambda+r)$ and $r\rightarrow\infty.$

\section{Asymptotic behaviour of the MLE}\label{Asym}

\subsection{Tightness and global asymptotic results}\label{sec:tight}

The first task in establishing asymptotic results is to show that the random variables in question are bounded in probability.
The following result does this for the well-specified setting.   The proof is quite technical and makes repeated use of log-concavity as well as characterization properties of the MLE.    These properties provide the necessary bounds in terms of the empirical distribution, from which tightness may be concluded.  We say that a pmf has one-sided support if this support can be written as $[\kappa, \infty)\cap\ZZ$ or $(-\infty, \kappa]\cap\ZZ$ for some $\kappa$ such that $|\kappa|<\infty.$

\begin{proposition}\label{BoundProb}
Suppose that $p_0$ is log-concave and has one-sided support, and let $r<s$ be two successive knots of $\psi_0.$  Then, for all $x \in \{r, \ldots, s-1\}$, $\sqrt n (\lmle_{n}(x) - \psi_{0}(x))$ and $\sqrt n (\mle_{n}(x) - p_{0}(x))$ are bounded in probability for all $n$ sufficiently large.
\end{proposition}

The approach used to prove Proposition \ref{BoundProb} cannot be applied to the misspecified setting.   We therefore use empirical process theory techniques to obtain the following global convergence rates.  Hellinger consistency of nonparametric maximum likelihood estimators was considered in \citet{geer93}, and these methods were later extended to the setting of misspecification in \citet{patilea} and \citet[Lemma 10.14]{geer}.    In both of these cases the underlying class of densities was convex, and therefore the cited results do not apply to our setting.    In \cite{BJRProofs2012}, we show how the methods of \cite{geer} and \cite{patilea} can be adapted to the class $\mc LC_1$.    The argument hinges on a new ``basic inequality" (see \citealp[Lemma D.2]{BJRProofs2012}),  which yields the following theorem.

\begin{theorem}\label{thm:rate}
Suppose that the support of $p_0$ is bounded.  Then $h(\widehat p_n, \widehat p_0)=O_p(n^{-1/2}).$
\end{theorem}
Convex classes of functions are ``easier" to handle since the basic inequality there (\citealp[Lemma 10.14]{geer} and \citealp[Lemma 2.2]{patilea}) gives bounds in terms of an empirical process on classes of bounded functions.  This is not the case in \citet[Lemma D.2]{BJRProofs2012}, and is the main reason that the support in Theorem \ref{thm:rate} is assumed to be bounded.   Because of this, it is not straightforward to extend our methods to the case of infinite support.    In fact, we believe that the (Hellinger distance) convergence rate in this setting will be slower than $\sqrt{n}.$  Hellinger distance is a strong metric for infinite sequences, so such a result would not be surprising.   For example, the empirical distribution $\emp_n$ is well-understood to converge at rates $\sqrt{n}$, both pointwise and in the $\ell_k$ sense (for $k\geq 2).$    However, $\emp_n$ converges at rate $\sqrt{n}$ in the Hellinger metric only for finite support (see, for example, \citealp[Corollary 4.3, Remark 4.4]{jankowski_09}).

\subsection{Pointwise asymptotics in the well- and misspecified settings}\label{sec:pointwise}

Fix a point $z$ which lies between two successive knots $r \leq z < s$ of $\widehat \psi_0$,  that are a finite distance apart.  The asymptotic distribution of the log-concave MLE at $z$ in both the well- and misspecified settings is described by the solution of similar least squares problems, and these are both defined below.   However, to establish the results, we need tightness to hold first, and therefore the assumptions used in our main result (Theorem \ref{thm:pointwise} below) are different in the well-specified and misspecified case.

In order to define the asymptotic distribution, we first require definitions of certain processes.  On the set $\{r, \ldots, s-1\},$ define  $\What_n(z) = \sqrt{n}(\widehat p_{n}(z)-\widehat p_{0}(z))$ and $\WW_n(z) = \sqrt{n}(\emp_{n}(z)-p_{0}(z)).$  Next, for $x\in \{r, \ldots, s\},$ let
\begin{eqnarray*}
\Hhat_n(x) = \sum_{y=r}^{x-1}\sum_{z=r}^y \What_n(z) \ \ \ \ \ \ \ \ & & \ \YY_n(x) = \sum_{y=r}^{x-1}\sum_{z=r}^y \WW_n(z),
\end{eqnarray*}
with the convention that $\Hhat_n(r)=\YY_n(r)=0.$  It is well-known that the processes $\WW_n$ and  $\YY_n$ have Gaussian limits.   Let $\WW(z)=\UU(F_0(z))-\UU(F_0(z-1)),$ where $\UU$ denotes a Brownian bridge from $(0, 0)$ to $(1, 0),$ and define the process $\YY(x) = \sum_{y=r}^{x-1} \sum_{z=r}^y \WW(z).$

We define the least squares (LS) functional
\bean\label{Phi}
\Phi(g) &=& \sum_{z=r}^{s-1} \ \widehat p_{0}(z) \left(g(z) - \frac{\WW(z)}{\widehat p_{0}(z)}\right)^2.
\eean
Next, let $\mc E=\{r \leq x <s: \sum_{y=r}^{x-1} (\widehat F_0(y)-F_0(y))=0 \},$
and define $\mc C|_{\mc E}$ as the class of concave functions on $r\leq z<s$ such that knots are only allowed in $\mc E$ (by definition, $r\in \mc E$).  Note that any element of this class satisfies
\begin{eqnarray}\label{line:char_miss}
g(z) &=& a+ b z + \sum_{y\in \mc E} c_y (y-z)_+,
\end{eqnarray}
where $a$ and $b$ are constants and $c_y\leq 0.$  It follows that the class $\mc C|_{\mc E}$ is convex.   Throughout we take the convention that $\sum_{y=r}^{r-1} h(y)=0$ for any choice of function $h$. The next result shows that a unique solution to \eqref{Phi} exists.  It also characterises its form.


\begin{proposition}\label{LS}
The functional $\Phi$ in \eqref{Phi} admits a unique minimiser, $g^*,$ over the class $\mc C|_{\mc E}$. Furthermore, $g =g^*$ if and only if
\begin{eqnarray*}
 \sum_{y=r}^{x-1} \sum_{z=r}^y g(z) \, \widehat p_{0}(z)
\left\{ \begin{array}{ll}
\leq& \YY(x), \ \ \forall \ x \in \mc E \\\\
= & \YY(x), \ \ \textrm{if $x \in \mc E $ is a knot of $g,$ or if x = s.}
\end{array}
\right.
\end{eqnarray*}
\end{proposition}
Note that if $p_0$ is log-concave, then $\widehat p_0 \equiv p_0$ and we have that $\mc E=\{r\leq x < s, x\in \ZZ\}$ and the class $\mc C|_{\mc E}$ is just the class of concave functions on $[r, s) \cap \ZZ.$  Let
\begin{eqnarray}\label{line:defHY}
\HH(x) = \sum_{y=r}^{x-1} \sum_{z=r}^y g^*(z) \, \widehat p_{0}(z)
\end{eqnarray}
for $x \in \{r, \ldots, s \}$. We are now able to state our main asymptotic result, taking the additional convention that $\HH(r-1)=\YY(r-1).$   For any function $f,$ let  $[\nabla f](x)=f(x+1)-f(x)$ denote the discrete gradient.

\begin{figure}[h!]
\begin{center}
\setkeys{Gin}{width=\textwidth}
\centerline{\includegraphics[width = 0.5\textwidth]{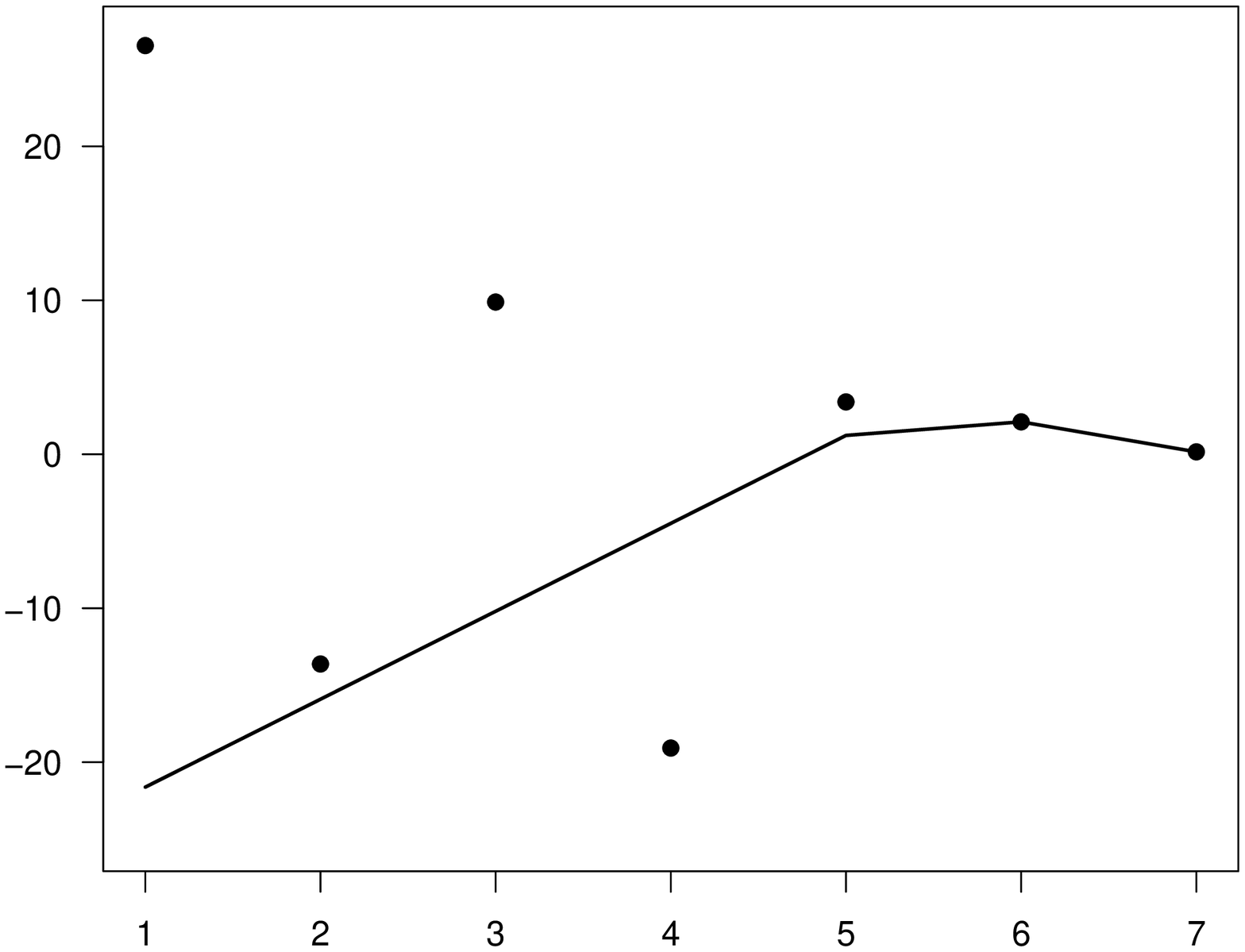}
\includegraphics[width = 0.5\textwidth]{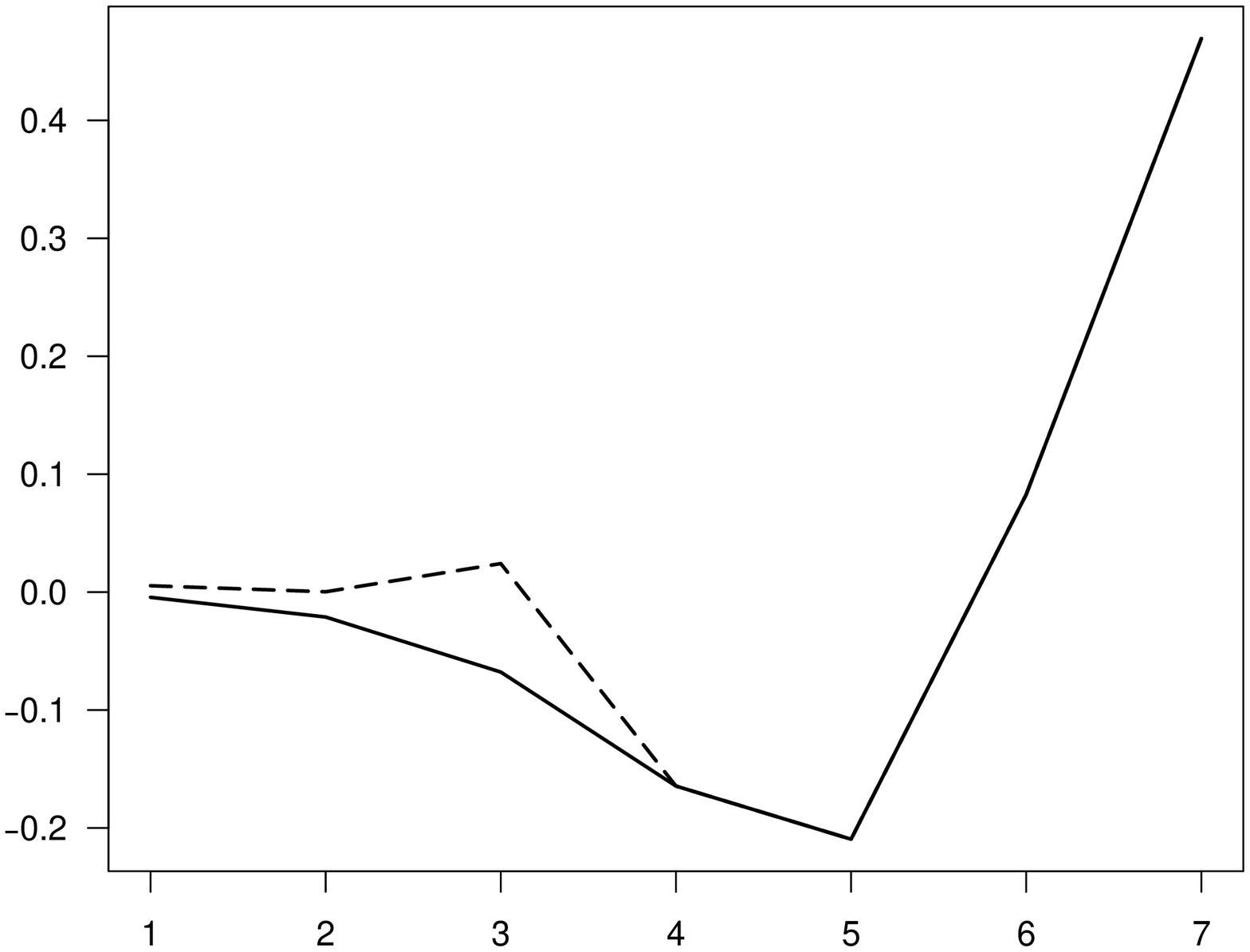}}
\end{center}
\vspace{-0.5cm}
\caption{The left plot shows $g^*$ (solid line) along with sampled values of $\WW(x)/p_{0}(x)$ (points) for a log-concave pmf (see \eqref{line:triangular} where $d=11$) with $\{r, \ldots, s-1\}=\{1, \ldots, 7\}$.  The right plot shows the corresponding processes $\HH(x)$ (solid) and $\YY(x)$ (dashed).}
\label{fig:LSsim}
\end{figure}

\begin{remark}\label{rem:sim}
Finding the minimiser of \eqref{Phi} is a weighted least squares concave regression problem.  For the case $\mc E=\{r, \ldots, s-1\}$ (the well-specified case), \eqref{Phi} can be minimised numerically using the function \normalfont{\code{conreg}} \textit{in the \proglang{R} package \pkg{cobs} \citep{cobs}. In this case the distribution of $(\WW(r), \ldots, \WW(s-1))$ is multivariate normal with zero mean and covariance matrix $(p_{0}(x) \delta_{x,y} - p_{0}(x)p_{0}(y))_{x,y = r, \ldots, s - 1}$, where $\delta_{x,y}$ denotes the Kronecker delta.  An example is shown in Figure \ref{fig:LSsim}. }
\end{remark}

In the following theorem, we describe the pointwise asymptotic behaviour of the log-concave MLE at a point $x$ in both well- and misspecified settings, and the assumptions differ between the two settings:
\begin{list}{--}
        {\setlength{\topsep}{3pt}
        \setlength{\parskip}{0pt}
        \setlength{\partopsep}{0pt}
        \setlength{\parsep}{0pt}
        \setlength{\itemsep}{3pt}
        \setlength{\leftmargin}{10pt}}
\item If $p_0$ is well-specified, then we assume that it has a (possibly infinite) one-sided support.  We also assume that $x$ lies between two knots which are a finite distance apart.   Note that in this case we have that $\widehat p_0=p_0.$
\item If $p_0$ is misspecified, then we assume that it has bounded support.
\end{list}

\medskip

\noindent Let $-\infty < r <s < \infty$ be two successive knots of $\widehat \psi_0$ (equal to $\psi_0$ in the well-specified case). If $s$ is not an internal knot of $\widehat \psi_0,$ then we set $s=s+1.$

\begin{theorem}\label{thm:pointwise}
Let $\HH$ denote the (unique) process on $\{r, \ldots, s\}$ as defined in \eqref{line:defHY}.   Then, for any $r\leq x<s$
\begin{eqnarray*}
\sqrt{n}(\mle_{n}(x)-\widehat p_{0}(x)) \stackrel{d}{\rightarrow} [\Delta \HH](x),  \ \ \sqrt{n}(\lmle_{n}(x)-\widehat \psi_{0}(x)) \stackrel{d}{\rightarrow} \frac{[\Delta \HH](x)}{\widehat p_{0}(x)}.
\end{eqnarray*}
Furthermore, if $p_0$ is log-concave, then
\begin{eqnarray*}
\sqrt{n}(\widehat F_n(x)-F_0(x)) \stackrel{d}{\rightarrow} [\nabla \HH](x) + \UU(F_0(r-1)).
\end{eqnarray*}
\end{theorem}
The proof of this theorem is relatively straightforward once we obtain tightness.   The characterization from Lemma \ref{CharMLE} is appropriately re-scaled, and, in the limit, this becomes the characterization of the LS process \eqref{line:defHY}.   Some additional care must be taken in the misspecified case.




\begin{remark}\label{AsympDoubleTripleKnot}
Recall the definition of a double and triple knot  given in Definition \ref{def:knot}. In the well-specified case, it follows from the above result that, when $r$ is a double knot, there is asymptotic equivalence between the log-concave MLE and the empirical pmf in the sense that the limiting distribution of $\sqrt{n}(\mle_{n}(r)-p_{0}(r))$  is the same as the limiting distribution of $\sqrt{n}(\emp_{n}(r)-p_{0}(r)) \stackrel{d}{\rightarrow} N(0, p_{0}(r)(1- p_{0}(r))).$ If $r$ is a triple knot, then it follows from Lemma \ref{CharMLE} that there exists $n_0$ such that
$$\mle_{n}(r) = \emp_{n}(r) \ $$
almost surely for all $n\geq n_0$. Hence, on any finite subset of the support of a strictly log-concave pmf, the log-concave MLE is almost surely equivalent to the empirical pmf, provided that $n$ is large enough. Examples of strictly log-concave pmfs include the binomial, negative binomial and Poisson distributions.
\end{remark}


In the correctly specified setting, the situation for the discrete case shares strong similarities with the one initially encountered by \cite{gjw01B} in convex estimation and afterwards in \cite{balabdaoui_09} in log-concave estimation in the continuous setting. In both works, the limit distribution of the nonparametric estimators involve a stochastic process that stays above (invelope) or below (envelope) a certain Gaussian process, whose second derivative is convex (concave) and upon which depend the limit of the estimators.  Knots of this second derivative are touch points of the in-/envelope and the Gaussian process.  To make the comparison more direct, note that the LS functional \eqref{Phi} may equivalently be defined as
\begin{eqnarray*}
\Phi(g) &=& \frac{1}{2}\sum_{z=r}^{s-1} g^2(z) \, p_0(z) - \sum_{z=r}^{s-1} g(z) \WW(z)
\end{eqnarray*}
when $p_0$ is log-concave.

\subsection{Confidence intervals for $p_0$ in the well-specified setting}\label{CIs}

One key application of the asymptotic results described above is that they may be used to calculate pointwise confidence intervals for the true log-concave pmf.   Recall that for our theory to apply, we assume that the true log pmf has only finite intervals between knot points, thus excluding geometric-like distributions.  However, these ``degenerate" cases form only a small subset of the class of log-concave pmfs.   Below, we describe how to compute 95\% confidence intervals, but the method can be generalised easily to any other coverage.   Furthermore, we describe how to calculate the intervals over the entire length of the support of the MLE. A similar approach can be also used over a smaller subsegment.

\begin{figure}[t!]
\begin{center}
\setkeys{Gin}{width=\textwidth}
\centerline{\includegraphics[angle=-90, width = 0.5\textwidth]{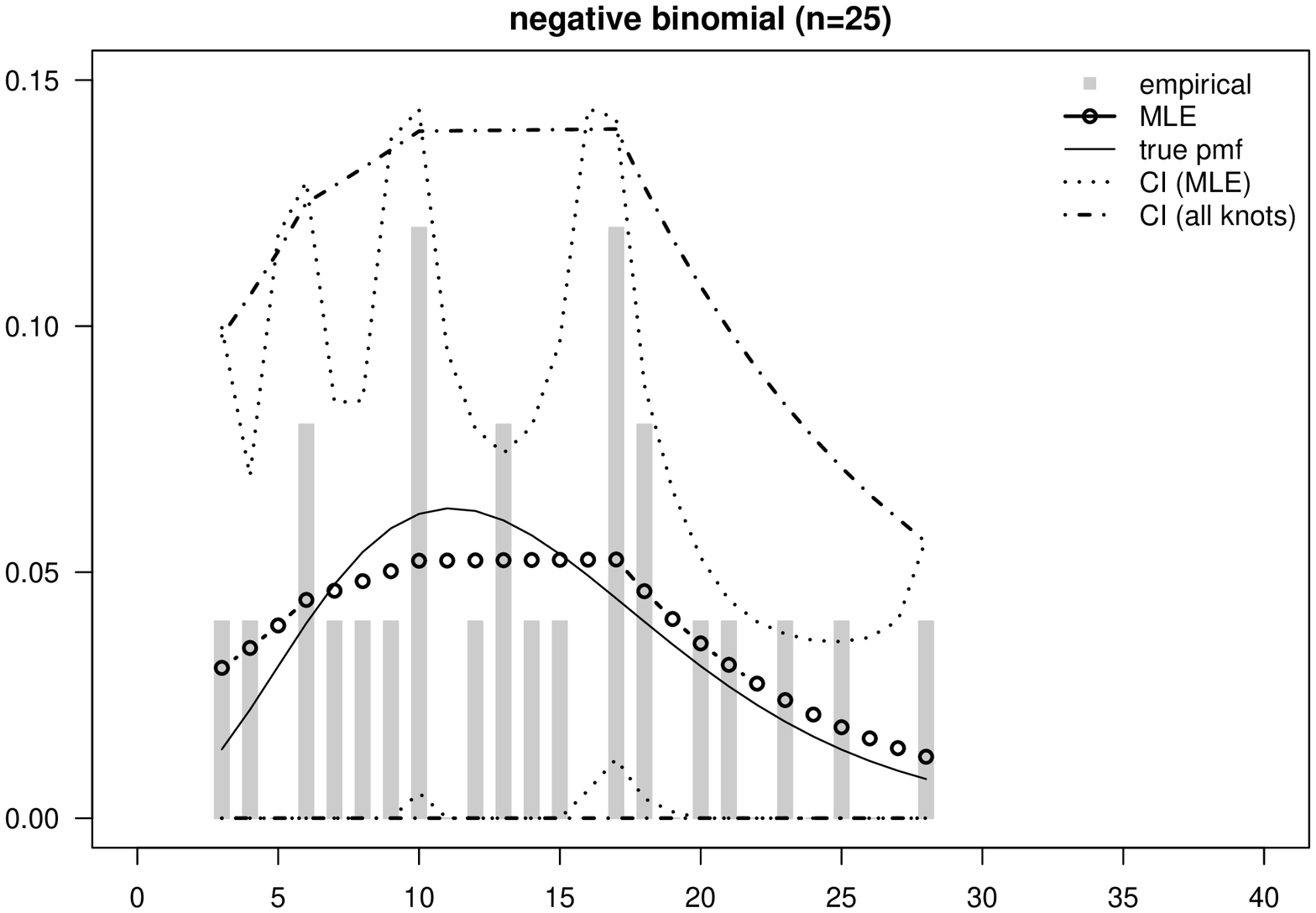}
\includegraphics[angle=-90, width = 0.5\textwidth]{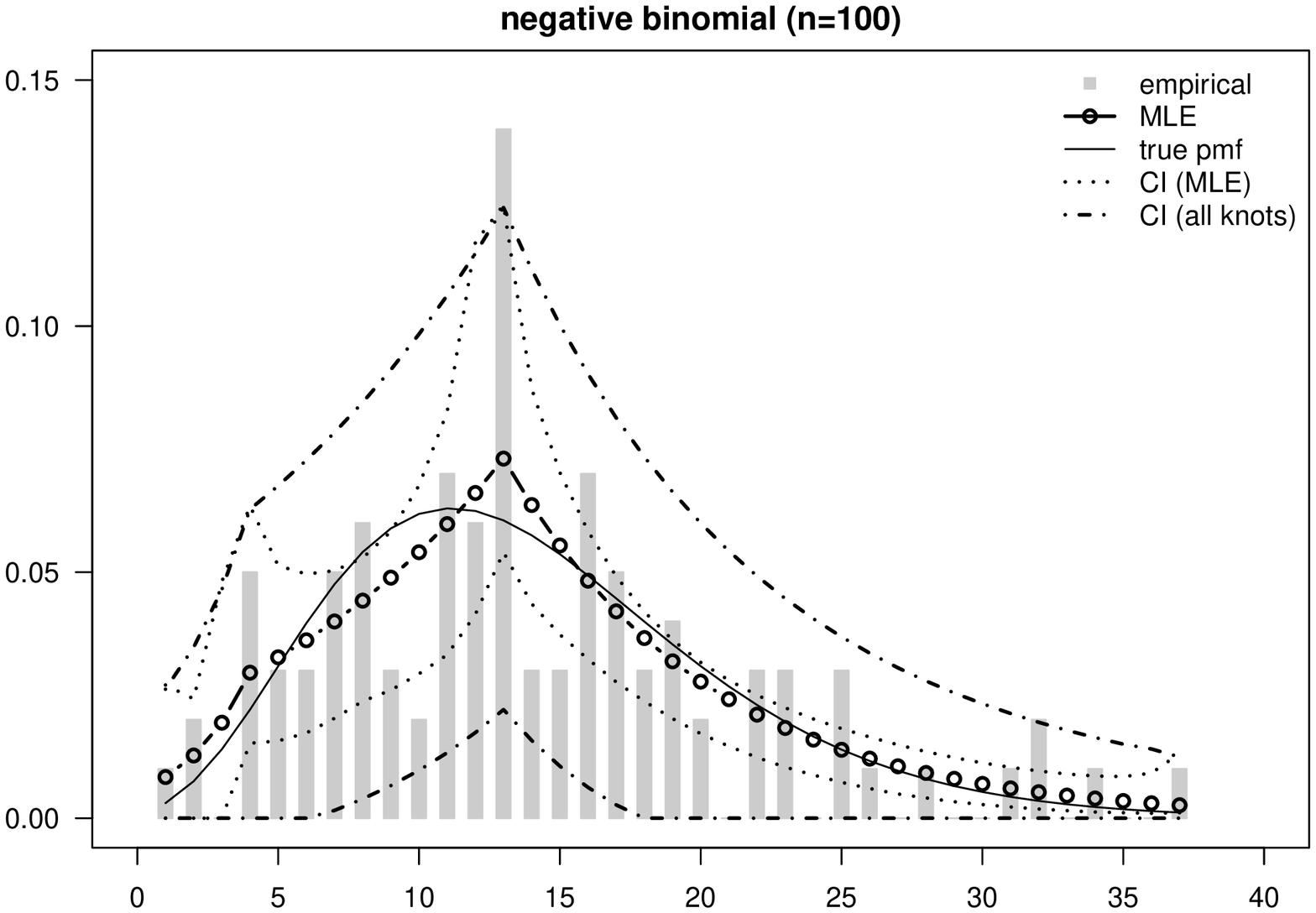}}
\end{center}
\vspace{-1cm}
\caption{Nonparametric MLE of the negative binomial (6, 0.3) distribution for $n=25$ (left) and $n=100$ (right).  Confidence intervals with knots based on the MLE (dotted line) and based on selection all points as knots (dash-dot line) are also shown.  As expected, the latter are wider.
}
\label{fig:CIintro}
\end{figure}


\begin{figure}[b!]
\begin{center}
\includegraphics[angle=-90, width = 0.45\textwidth]{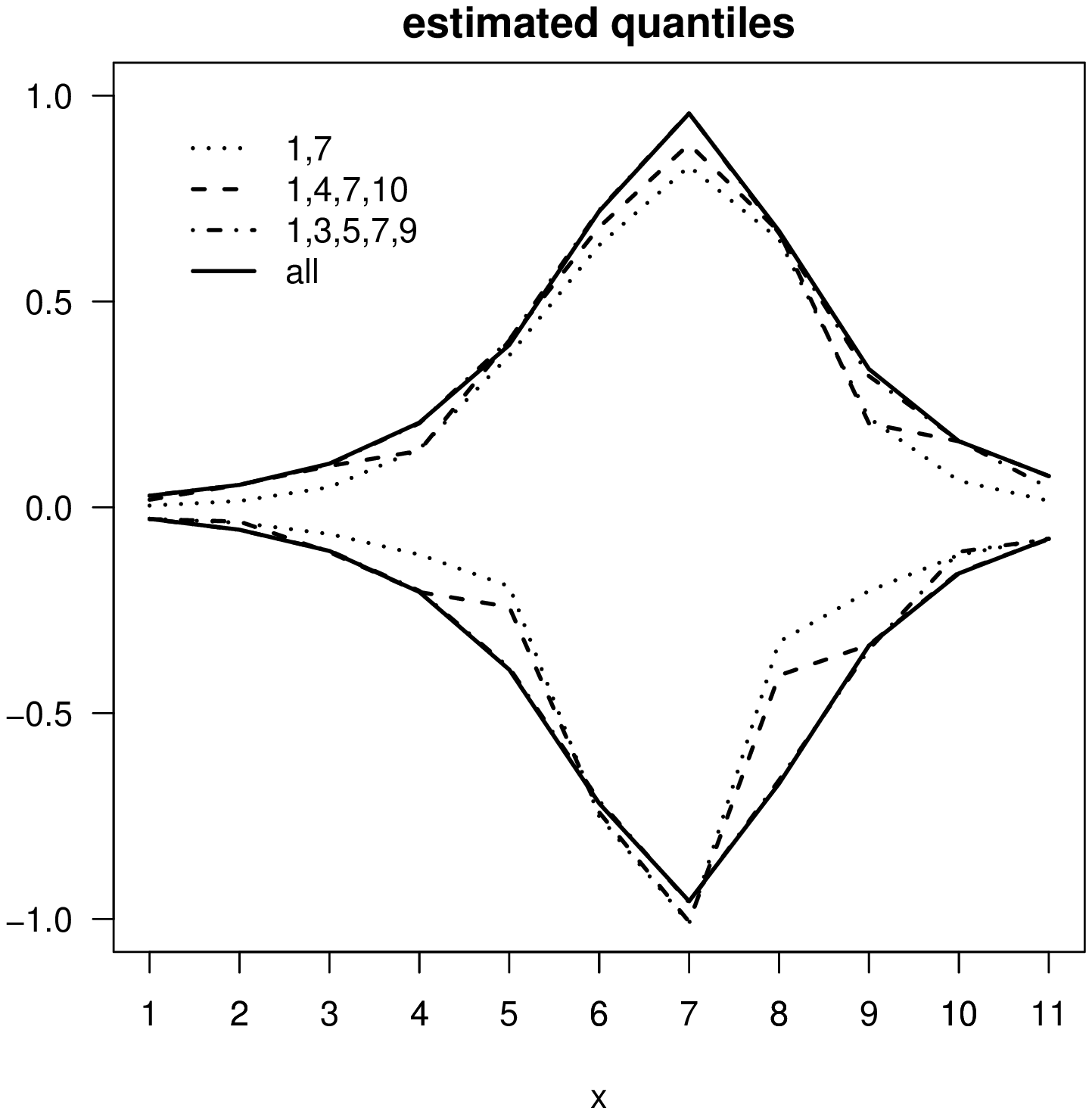}
\includegraphics[angle=-90, width = 0.45\textwidth]{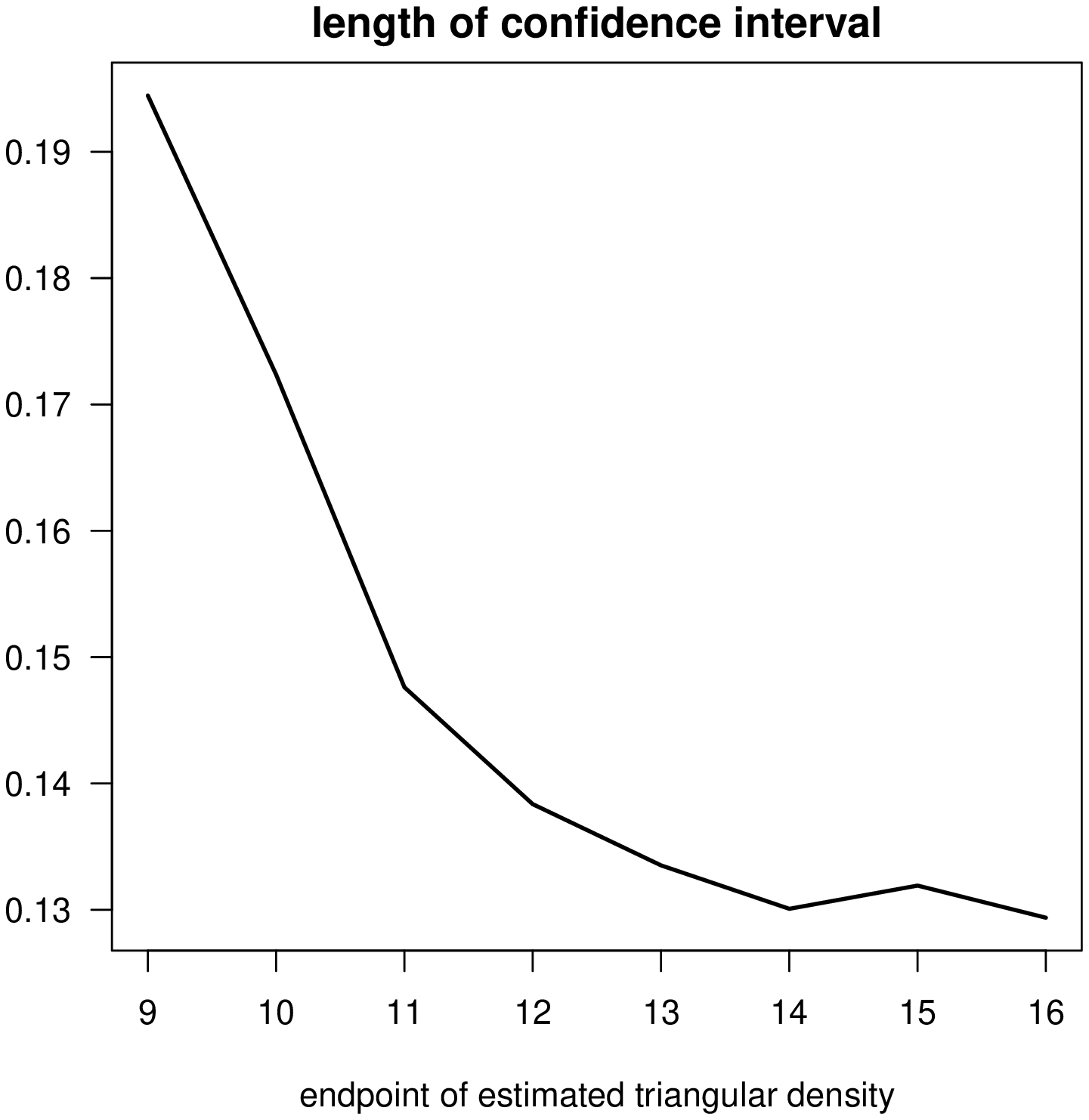}
\end{center}
\vspace{-0.5cm}
\caption{Estimated 95\% quantiles at different points (left) and lengths of confidence intervals at one point (right) for the triangular pmf \eqref{line:triangular}.
On the left, $d=11$ and the quantiles were estimated from the true pmf and assuming different knot points, as indicated.  {On the right, the lengths of the confidence intervals at $x=9$ for varying values of $a$ in \eqref{line:triangular} were estimated using the MLE.  The mean lengths are plotted against the endpoint of the empirical pmf.}}
\label{fig:Qtriangular}
\end{figure}

Let $\mc S$ denote the support of the true pmf, and write $\mc S= \cup_j I_j,$ where $I_j=\{r_j, \ldots, r_{j+1}-1\},$ where the $r_j$ denote the knot points of the true log-pmf.   For each $x\in I_j,$ let $q_1(x), q_2(x)$ denote 2.5\% and 97.5\% quantiles of the distribution of $[\Delta \HH](x).$   Since we can simulate directly from the distribution of $[\Delta \HH](x),$ these are straightforward to estimate.  Then,
\begin{eqnarray*}
\{\mle_{n}(x) - \widehat q_2(x)/\sqrt{n}, \mle_{n}(x) - \widehat q_1(x)/\sqrt{n}\},
\end{eqnarray*}
give approximate confidence intervals, which have asymptotically correct coverage.    Note that if $|I_j|=1,$ then for $x\in I_j, $  $q_1(x) = -1.96 \sqrt{p_{0}(x)(1-p_{0}(x))}$ and $q_2(x) = -q_1(x),$ by Remark~\ref{AsympDoubleTripleKnot}.

To estimate the quantiles $q_1$ and $q_2$ we need to estimate the true pmf, including the true knots of the log-pmf.  The true pmf is easily estimated by the MLE, but a more serious issue is that we do not know the true locations of the knots.  We propose to estimate these as the knots of the log-MLE $\lmle_n.$  As noted following Lemma \ref{lemma:knots}, the knots of $\lmle_n$ will, at worst, asymptotically overestimate the set of true knots.    We believe that the penalty for this is a slight overestimation of the quantiles.  Our reasoning relies on Remark~\ref{remark:DOF} and the following discussion. Overestimating the true set of knots causes us to overestimate the degrees of freedom of the estimator, which in turn means that we tend to overestimate the quantiles in the confidence intervals.   This idea is confirmed through simulation, as shown in Figure \ref{fig:Qtriangular}, where we consider quantiles and confidence interval lengths for the triangular pmf
{
\begin{eqnarray}\label{line:triangular}
p_{0}^a(x) &\propto& \left\{\begin{array}{ll}
					 \exp\{4(x-1)/3\} & \mbox{ for } x \in \{1, \ldots, 7\},\\
					 \exp\{(16-a)(x-7)/(a-7) +8\} & \mbox{ for } x \in \{8, \ldots, a\}.
				     \end{array}\right.	
\end{eqnarray}
To assess the association between the length of a linear stretch of a pmf and the length of our proposed confidence interval, we simulated $M = 500$ confidence intervals at $x=9$ for samples of size $n = 50$ and values of $a \in \{9, \ldots, 16\}.$  The number of re-sampling draws to compute the quantiles was $B = 1000$. The right plot in Figure~\ref{fig:Qtriangular} provides mean lengths of these confidence intervals and reveals that the longer a linear stretch gets, the shorter the average confidence interval length is.
}

\section{Analysis of H1N1 incubation and symptom durations}\label{Examples}

\begin{figure}[t!]
\begin{center}
\setkeys{Gin}{width=\textwidth}
\centerline{\includegraphics[angle=-90, width = 0.45\textwidth]{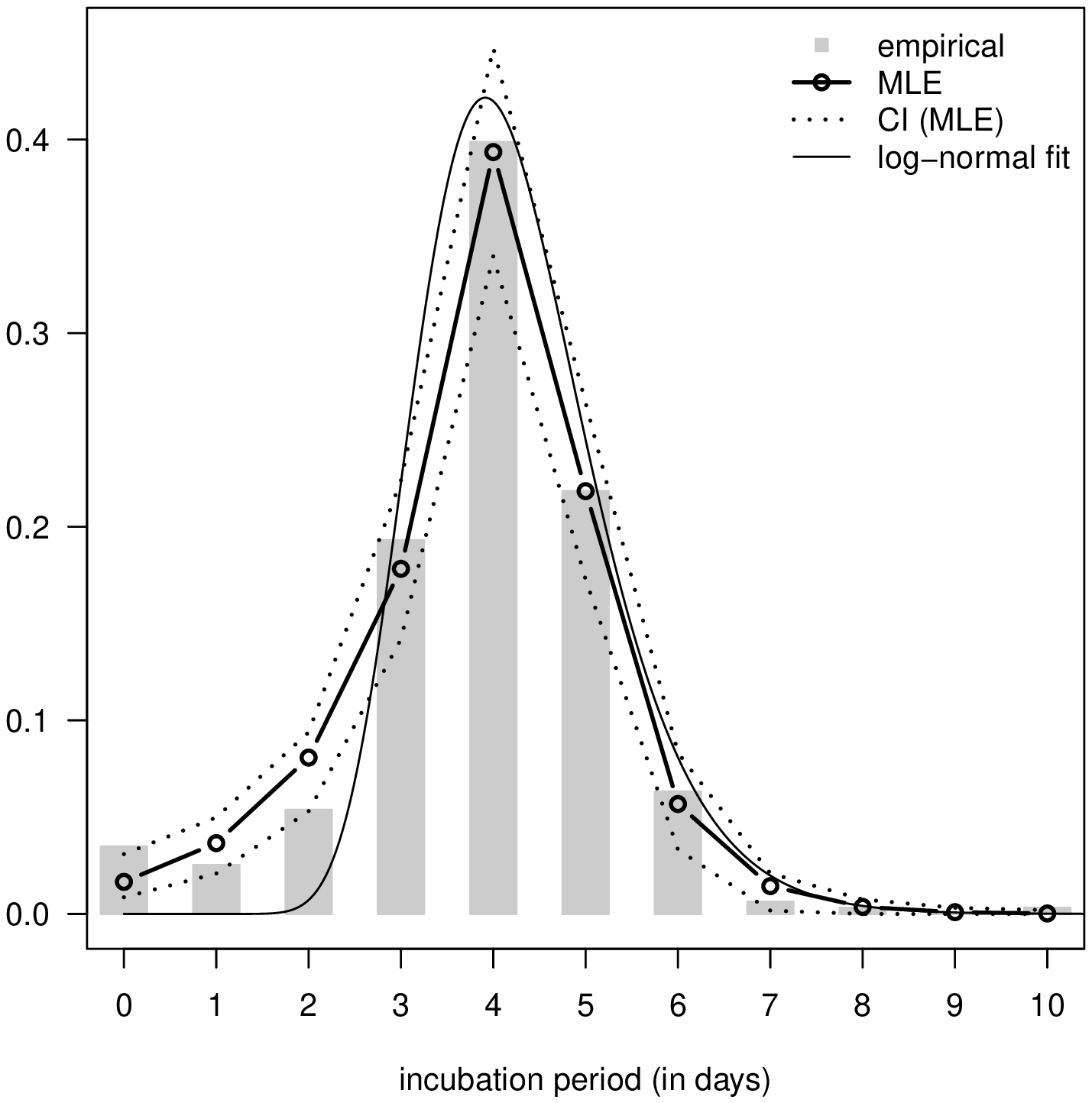}
\includegraphics[angle=-90, width = 0.45\textwidth]{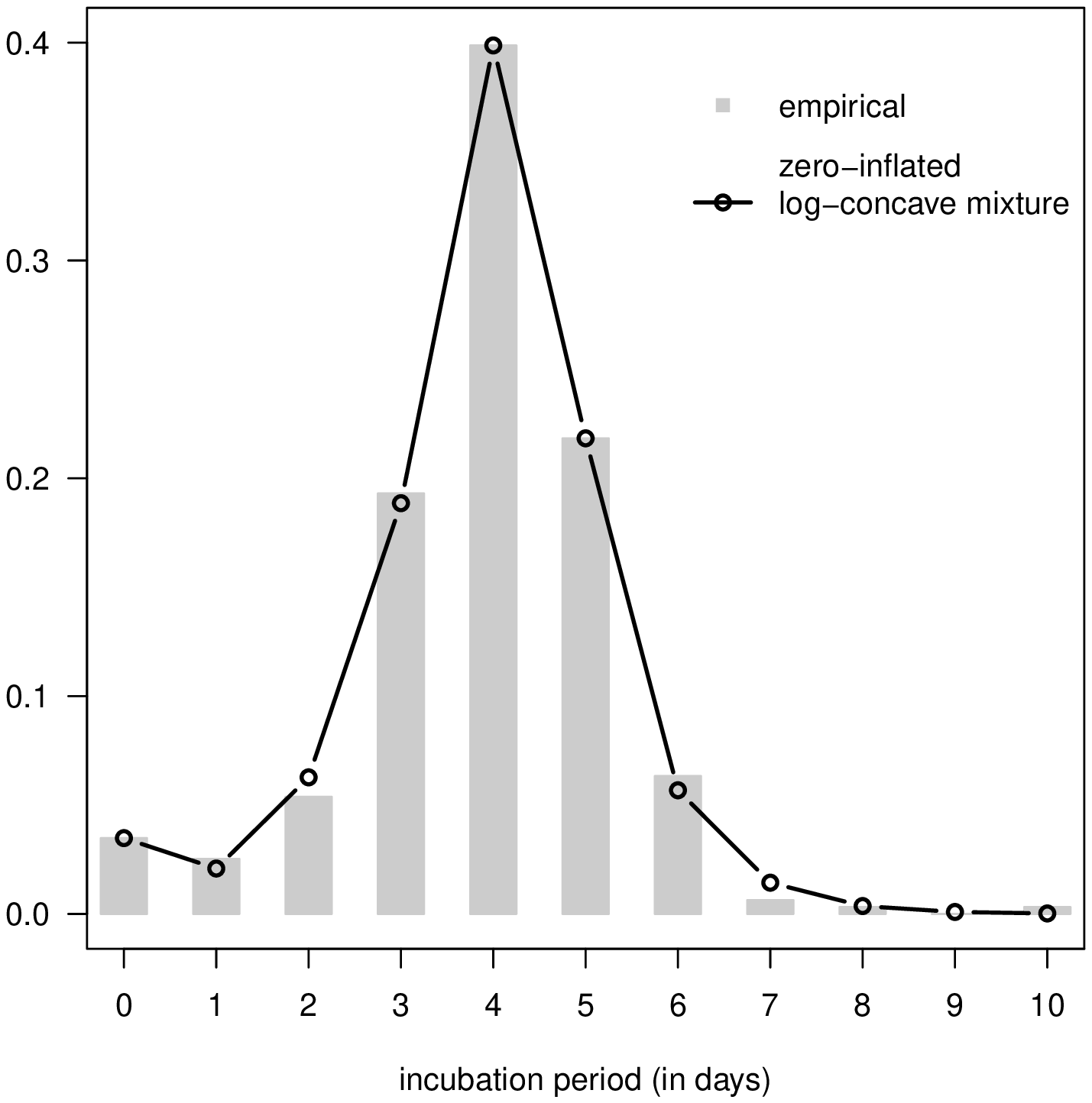}}
\end{center}
\vspace*{-1cm}
\caption{Estimates of incubation period for the H1N1 data.} \label{influenza:incub}
\end{figure}

We illustrate the new estimator on H1N1 influenza data from Canada.  \cite{tuite_10} report an early study of the H1N1 pandemic.  The goal of the study was to understand the behaviour of the disease, including the incubation period (time from exposure to the disease to onset of symptoms) and the duration of symptoms.   H1N1 individual-level data was collected for laboratory--confirmed cases of the disease for a 3-month period in the spring of 2009.   From these, information on the incubation period (in days, $n=316$) and symptom duration (in days, $n=712$) was derived.   For more details on data acquisition we refer to \citet{tuite_10}.

Clinicians and mathematical biologists are most often interested in the \emph{fitted} mean, standard deviation, and range to understand the behaviour of the virus.  These may be used in a sensitivity analysis of the developed deterministic and stochastic models, as was done in \citet{tuite_10}.   For example, output of these models would be checked against known behaviour from the fitted distributions, to ascertain the appropriateness of the former.   Alternatively, the model may use the fitted distribution itself within the algorithm, and hence requires the ability to simulate from the fitted distribution.

\begin{figure}[t!]
\begin{center}
\includegraphics[angle=-90, width = 0.95\textwidth]{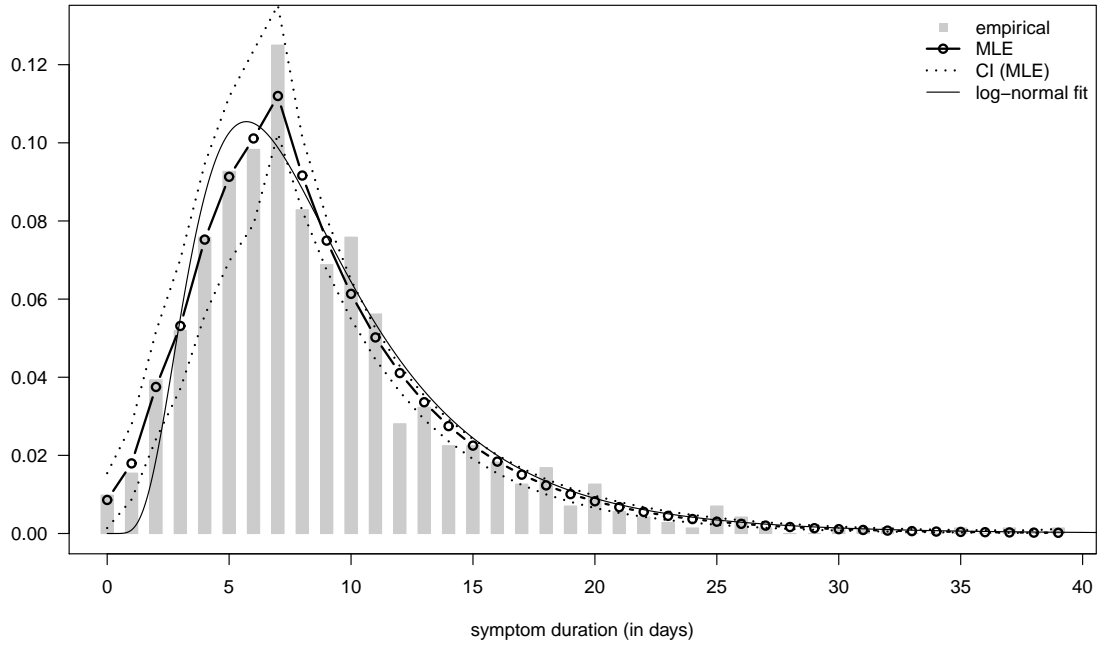}\\
\includegraphics[angle=-90, width = 0.95\textwidth]{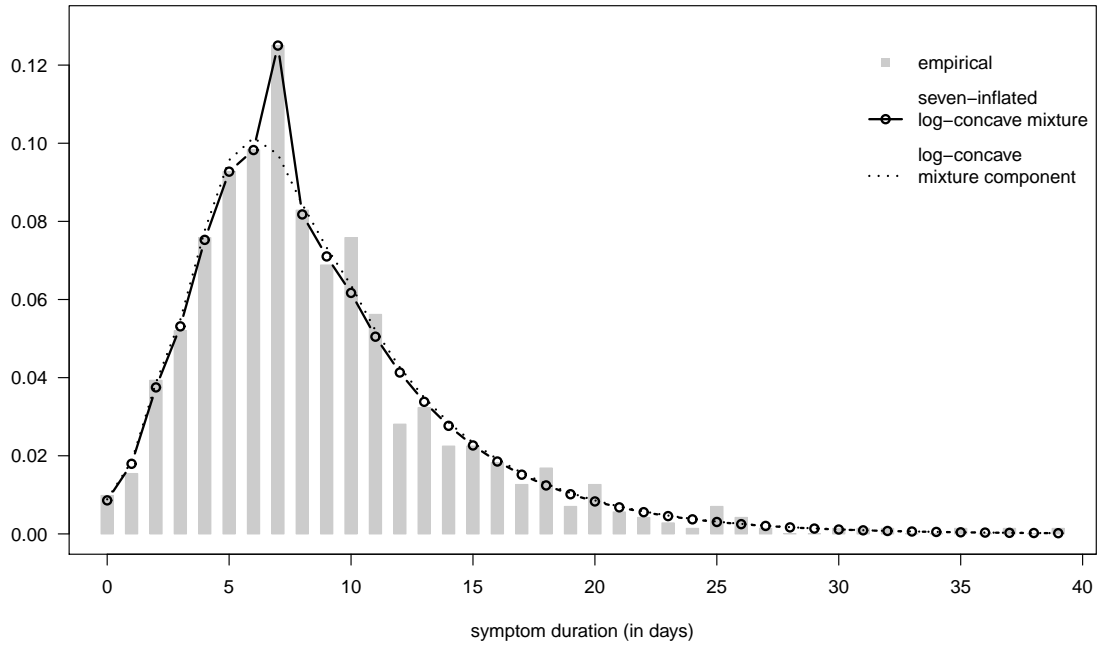}
\end{center}
\vspace*{-0.5cm}
\caption{Estimates of duration for H1N1 data.} \label{influenza:dur}
\end{figure}

In \cite{tuite_10}, a log-normal distribution and  Weibull distribution were fit to both data sets.  To estimate the densities the authors used Excel's \proglang{solver}. After assessing goodness-of-fit, the final model chosen was the log-normal distribution.    In Figure~\ref{influenza:incub} (left) and Figure~\ref{influenza:dur} (top), we show the log-normal fitted distributions and compare it with the log-concave MLE.  Pointwise 95\% confidence intervals based on the MLE are also shown.   It is easy to see that the log-normal does not capture well some key aspects of the empirical distribution.  We make the following notes about the log-concave MLE.

\begin{list}{--}
        {\setlength{\topsep}{3pt}
        \setlength{\parskip}{0pt}
        \setlength{\partopsep}{0pt}
        \setlength{\parsep}{0pt}
        \setlength{\itemsep}{3pt}
        \setlength{\leftmargin}{10pt}}
\item  Both the incubation data and duration data have been grouped, and therefore a discrete/grouped model is more appropriate than a continuous one.

\item The MLE captures well the shape of the empirical distribution, including the mode and the height of the mode.  Notably, the MLE has the same mean and range as the empirical distribution.  As shown in \eqref{line:ExpBound}, the MLE will have a smaller variance than the empirical distribution.

\item Having estimated the MLE it is very easy to sample from its distribution.  Given the seemingly accurate fit of our new nonparametric estimates compared to the empirical pmf, we argue that these random numbers would be more accurate than those from the log-normal model.

\item Log-concavity encompasses many parametric models, but is substantially more flexible than any particular parametric model and can capture a wide range of possible shapes.  Moreover, the MLE is fully automatic, as it does not necessitate a choice of kernel, bandwidth, or prior.   In this example, the MLE fits the empirical well, and it also  ``smooths'' the empirical especially in the rather variable tail of the symptom duration distribution.

\item In the analysis of an infectious disease, the incubation period is of great importance, particularly so in the lower tail of the distribution, as this provides information on the rate of spread of the virus within a population.  The log-normal does not fit well the lower tail of the empirical distribution, and the MLE is better at describing this behaviour.  A closer examination of the empirical data shows a spike at zero, which is most likely caused by inaccurate reporting of the onset of symptoms.  To better describe this behaviour, we also fit a mixture of a log-concave pmf with a point mass at zero.   This is, essentially, a zero-inflated log-concave distribution.   
The results are shown in Figure \ref{influenza:incub} (right).    
The mean of the pure MLE was 3.88, which is equal to the mean of the data.  The mean of the log--concave part of the mixture model was slightly higher, at 4.02.    

\item The data for the duration of symptoms of the swine flu has a clear spike at $t=7$ days.    As above, this is probably caused by mis-reporting, as seven days is equivalent to one week, and therefore a likely choice in a patient's response.   One ad-hoc method to account for this, is to again fit an inflated model, this time placing the point mass function at $t=7$ days.  The results are shown in Figure \ref{influenza:dur} (bottom).    
The mean of the pure MLE was 8.66, which was also the mean of the fitted mixture model.  The mean of the log-concave part of the mixture model is slightly higher, at 8.72, however, the log-concave component has a lower mode at $t=6$.    The probability of observing an ``inflated" value at seven was found to be~0.031.  

\item In addition to the aforementioned issue, it is quite likely that the duration data collected suffers from length-bias (see e.g. \citealp{mamoud:00}), in that those with longer duration of symptoms were more likely to be observed.   It would be of interest to see if our methods can be modified to include a length-bias correction, however, this is beyond the scope of this work.  

\end{list}

\section{Discussion}\label{conclusion}

In this paper we have studied estimation of a log-concave probability mass function via the nonparametric MLE.   Our simulations show that, for finite sample size, the log-concave MLE has behaviour superior to that of the empirical pmf, and can even be competitive with the parametric MLE (see the negative binomial example in Figure \ref{fig:smallsample}).

The main theoretical results of this paper establish explicitly the limiting distribution of the maximum likelihood estimator at a point of a log-concave MLE in the well- and misspecified settings.  In both cases, the limiting distribution is given in terms of a least squares problem and can also be described in terms of an envelope-type process (cf. \citealp{gjw01A, gjw01B, balabdaoui_09}).  In the well-specified case, the R package \pkg{cobs} allows us to solve the least squares problem.  This property was exploited in Section \ref{CIs}, where confidence intervals for the well-specified $p_0$ were developed.

Our results show that the pointwise convergence in the misspecified setting is of rate $\sqrt{n},$ and we identify the limiting distribution.  The importance of such understanding is clear:  it is the first step in assessing the power in the hypothesis testing of log-concavity.   For example, suppose that the hypothesis test is based on some functional $T(\mle_n)$.  Our results indicate that, at least in the case of bounded support, the power will depend on the size of $\sqrt{n}(T(\widehat p_0)-T(p_0)),$ as expected.  Various hypothesis testing methods have been considered in \citet{an, cule_08, hall05, hazelton_11, walther_02}.  Extending our results to handle global convergence rates when the support is unbounded is of considerable interest, but this problem requires the development of additional techniques, as the methods employed in \citet{jankowski_09} do not immediately apply to the log-concave pmf setting.

\section*{Acknowledgments}
We would like to thank Filippo Santambrogio and Jon Wellner for helpful discussions and Kathrin Weyermann and Lutz D\"umbgen for sending us a copy of Kathrin's Master's thesis and the \proglang{Matlab} code to compute the MLE. We owe many thanks to Ashleigh Tuite and David Fisman who made the H1N1 data available to us, and Jane Heffernan who provided us with much insight into the data. Finally, we would like to thank the AE and two anonymous referees for their invaluable comments and suggestions, all of which greatly contributed to improving the original manuscript.

\bibliographystyle{ims}
\bibliography{logconcave}

\end{document}